\documentstyle[epsf,preprint,aps]{revtex}
\begin{document}
\tighten
\draft
\preprint{ }
\title{HAMILTONIANS FOR REDUCED GRAVITY}
\author{James D. E. Grant and Ian G. Moss}
\address{
Department of Physics, University of Newcastle Upon Tyne, NE1 7RU U.K.
}
\date{March 1997}
\maketitle
\begin{abstract}
A generalised canonical formulation of gravity is devised for foliations of spacetime with codimension $n\ge1$. The new formalism retains $n$-dimensional covariance and is especially suited to $2+2$ decompositions of spacetime. It is also possible to use the generalised formalism to obtain boundary contributions to the $3+1$ Hamiltonian. 
\end{abstract}
\pacs{Pacs numbers: 04.20.Fy, 04.70.Dy}
\narrowtext
\section{INTRODUCTION}

The canonical methods introduced into the theory of General Relativity by Arnowitt, Deser and Misner\cite{arnie} involve a certain point of view in which spacetime is decomposed into spacelike hypersurfaces of constant time. However, there are some situations were a different decomposition of spacetime may be prefered and for these situations a new form of canonical method is required.

An important example is provided by the $2+2$ decompositions of General Relativity\cite{geroch,dinverno,hayward,brady} that arise naturally in the characteristic initial value problem for the Einstein equations. Constructing the Hamiltonian along conventional lines requires singling out one of the two directions\cite{torre}. It would be desirable to retain some advantages of the Hamiltonian formalism without breaking the two-dimensional covariance.

Another situation of interest arises from considering the thermodynamic properties of black holes\cite{hawking}. In the usual setup, one envisages a black hole and thermal radiation in equilibrium inside a box. The geometry on the surface of the box is fixed when evaluating the partition function\cite{york1,york2,moss}. This introduces boundary terms into the Hamiltonian. The origin of these boundary terms can be seen in a $2+2$ decomposition close to the boundary\cite{york3,brown,hayward1,hayward2,hawking2}.

We will describe a generalised canonical method that is appropriate to these situations. In the standard canonical formalism, time is introduced by foliating spacetime with spacelike hypersurfaces. We will base our canonical formalism on a foliation of codimension $n$, where $n\le d$, the dimension of spacetime. This gives us $n$ coordinates $x^I$, not all of them timelike.

For each variable $q^i$ we define $n$ momenta $p^I_{\ i}$ in terms of the Lagrangian density $L(q,\partial_Iq,\partial_aq)$, where $\partial_I$ are normal derivatives and $\partial_a$ tangential derivatives. We set
\begin{equation}  
p^I_{\ i}={\partial L\over \partial(\partial_Iq^i)}
\end{equation}
and take the definition of the Hamiltonian density to be
\begin{equation}
H=p^I_{\ i}\partial_Iq^i-L
\end{equation}
A Legendre transformation can be used to express the field equations obtained from the Lagrangian in Hamiltonian form
\begin{eqnarray}
\partial_Iq^i&=&{\delta H\over\delta p^I_{\ i}}\\
\partial_Ip^I_{\ i}&=&{\delta H\over \delta q^i}.
\end{eqnarray}
These equations can also be obtained by independent variation of the variables and the momenta in the action.

The case $n=d$ has a history stretching back to De Donder\cite{dedonder} and Weyl\cite{weyl}. This has been extended to General Relativity comparatively recently\cite{horova}. The geometrical structure of the Weyl--De Donder theory has also been investigated\cite{gotay,kan}. We will work within the range $1\le n<d$, intermediate between the usual canonical approach and the Weyl-De Donder theory.

The most significant advantage to be gained by introducing a Hamiltonian is the existance of canonical transformations. These exist also for the generalised formalism\cite{gotay,kan} and this gives us the freedom to perform changes of variables that simplify the resulting field equations. An example of this is provided in section 4, where we consider the $2+2$ decomposition of the Einstein equations.

There are also important limitations to the method. The Hamiltonian density, integrated over a hypersurface, does not generate translations along the $x^I$ directions in the way that the usual Hamiltonian does. This complicates the generalisation of Hamilton--Jacobi theory and implies that the generalised Hamiltonian density does not have a direct application to quantum theory. 

A positive feature of the more general approach is that the momenta have a natural variational interpretation. In the usual canonical formalism, the variation of Hamilton's principal function with respect to the fields at the final time gives the momenta. In section 5 we see that, in the new approach, the momenta can be obtained from variation of a principal function with respect to fields fixed on a corresponding boundary. This makes the new approach a natural way of analysing the partition function for black holes.

\section{LAGRANGIAN DECOMPOSITON}

We suppose that the original manifold ${\cal M}$ can be decomposed in $n<d$ coordinate surfaces, with coordinates $x^I$, and $m$ dimensional subspaces $\Sigma$. Our aim is to decompose the Einstein-Hilbert Lagrangian for the metric $\bf g$,
\begin{equation}
L={1\over2\kappa^2}R|g|^{1/2},
\end{equation}
where $\kappa^2=8\pi G$ and $|g|=|\det g_{ab}|$, into a kinetic term, which has derivatives with respect to $x^I$, and a potential term.

The decomposed metric $\bf g$ can be written
\begin{equation}
{\bf g} =  \sigma_{ab}\mbox{\boldmath$\omega$}^a \otimes \mbox{\boldmath$\omega$}^b +
{\eta}_{IJ} {\bf n}^I \otimes {\bf n}^J,
\end{equation} 
where $\mbox{\boldmath$\omega$}^a$ belongs to the cotangent space of $\Sigma$ and ${\bf n}^I$ to the orthogonal space. We choose a coordinate basis in which ${\bf n}^I$ and the dual vectors ${\bf n}_I$ are given by
\begin{equation}
{\bf n}^I = dx^I, \qquad
{\bf n}_I = {\partial}_I - N_I{}^a {\partial}_a,
\end{equation}
where the ${\partial}_a$ are tangential derivatives.

When a covariant expression is wanted we replace the coordinate derivative by the Lie derivatives ${\cal L}_I$ along ${\bf n}_I$. The tangential components of the Lie derivative will be denoted by $D$,
\begin{equation}
D_IT^{b\dots c}{}_{d\dots e}=
\sigma_{b'}^{\ b}\dots \sigma_{c'}^{\ c}\sigma_d^{\ d'}\dots \sigma_e^{\ e'}
{\cal L}_IT^{b'\dots c'}{}_{d'\dots e'}\label{ddef}
\end{equation}
The derivative $D$ is the same as the one used in references \cite{cho} and  \cite{brady}. We also define tangential vector fields
\begin{equation}
F_{IJ} = \left[ D_I,D_J\right],\label{f}
\end{equation}
which depend only on the $N_I^{\ a}$.

Decompositions of the curvature tensor are described in the appendix. These allow us to express the Ricci curvature scalar as
\begin{eqnarray}
R&=&{}^\|R+{}^\perp R+k_Ik^I-k_{Iab}k^{Iab}\nonumber\\
&&+a^I_{\ I}{}_aa^J_{\ J}{}^a-a^{IJa}a_{JIa}+\nabla_av^a.\label{ric}
\end{eqnarray}
where
\begin{eqnarray}
k^I_{\ ab}&=&\case1/2\eta^{IJ}D_J\sigma_{ab}\\
a^{IJ}{}_a&=&\case1/2F^{IJ}{}_a+\case1/2 {\partial}_a({\eta}^{IJ}).
\end{eqnarray}
Also,
\begin{equation}
v^a=-2n_I^{\ \,a}k^I+2a^I_{\ I}{}^a.\label{b1}
\end{equation}
contributes boundary terms.

The Ricci curvature scalar ${}^\perp R$ can also be expressed in terms of derivatives,
\begin{equation}
{}^\perp R=c^{IJM}c_{JIM}-c^{I\ \ M}_{\ \,I}c^J_{\ \ MJ}
-2c^I_{\ \ [IJ]}k^J-2a^{IJ}{}_aa_{[IJ]a}+\nabla_bu^b
\end{equation}
where
\begin{equation}
c_{IJK}=-\case1/2D_I\eta_{JK}-
\case1/2D_K\eta_{IJ}+\case1/2D_J\eta_{IK}
\end{equation}
and
\begin{equation}
u^b=\eta^{IK}(D_I\eta_{JK}-D_J\eta_{IK})n^{Jb}.\label{b2}
\end{equation}
contributes boundary terms.

Putting all this into the Einstein-Hilbert Lagrangian, we get
\begin{equation}
L=T-V
\end{equation}
where
\begin{eqnarray}
T&=&{1\over 2}G^{(ab)(cd)}\eta^{IJ}D_I\sigma_{ab}D_J\sigma_{cd}
+{1\over 2}G^{(IJ)K(LM)N}D_K\eta_{IJ}D_N\eta_{LM}
+G^{(IJ)(KL)}\sigma^{ab}D_K\eta_{IJ}D_L\sigma_{ab}\nonumber\\
&&-{1\over 8\kappa^2}|g|^{1/2}\sigma_{ab}F_{IJ}{}^aF^{IJb}\label{td}\\
V&=&-{1\over 2}G^{(IJ)(KL)}\sigma^{ab}\partial_a\eta_{IJ}\partial_b\eta_{KL}
-{1\over 2\kappa^2}|g|^{1/2}{}^\|R.\label{pot}
\end{eqnarray}
The tensors appearing in the kinetic term are
\begin{eqnarray}
G^{abcd}&=&{1\over 4\kappa^2}|g|^{1/2}\left(
\sigma^{ab}\sigma^{cd}-\sigma^{ac}\sigma^{bd}\right)\nonumber\\
G^{IJKLMN}&=&{1\over 2\kappa^2}|g|^{1/2}\left(\eta^{IL}\eta^{JN}\eta^{KM}
+\eta^{I[J}\eta^{L]M}\eta^{KN}
+\eta^{IJ}\eta^{MN}\eta^{KL}\right)\nonumber\\
G^{IJKL}&=&{1\over 4\kappa^2}|g|^{1/2}
\left(\eta^{IJ}\eta^{KL}-\eta^{IL}\eta^{JK}\right)
\end{eqnarray}
{\it symmetrised} as indicated by brackets in equation (\ref{td}).

Similar results to this have been obtained before. The case $n=2$ has been analysed by Torre\cite{torre}. Results for general $n$ are quoted by Cho et al. \cite{cho}, although their Lagrangian still retains terms that are second order in derivatives.

\section{HAMILTONIAN DECOMPOSITION}

The Hamiltonian formalism that we develop here is a generalisation of the usual approach, with special attention to retaining the covariance along the tangential and normal directions.  Momenta are defined in each of the normal directions $x^I$ conjugate to the quantities $\sigma_{ab}, {\eta}_{IJ}$ 
and $N_I{}^a$ as 
\begin{eqnarray}
\pi^{IJK} &=& 
{{\delta}\over{\delta ({\partial}_K {\eta}_{IJ})}} S,\label{moma}
\\
 p^{I\,ab} &=&
{{\delta}\over{\delta ({\partial}_I \sigma_{ab})}} S,
\\
P^{IJ}{}_a &=& 
{{\delta}\over{\delta ({\partial}_I N_J{}^a)}} S.\label{momc}
\end{eqnarray}

We shall see shortly that these relations can be inverted when the codimension $n>1$, enabling us to replace derivatives of fields along the normal directions with momenta. The case $n=1$ leads to the usual Hamiltonian formalism with constraints.

We define our Hamiltonian density to be
\begin{equation}
H = \pi^{IJK} {\partial}_K {\eta}_{IJ} 
+ p^{I\,ab} {\partial}_I \sigma_{ab} 
+ P^{IJ}{}_a {\partial}_I N_J{}^a - L.
\end{equation}
This can be written in covariant form using the derivative $D$,
\begin{equation}
H=\pi^{IJK} D_K {\eta}_{IJ} 
+ p^{I\,ab} D_I \sigma_{ab} 
+\case1/2 P^{IJ}{}_a F_{IJ}{}^a +N_I^{\ a}H^I_{\ a}- L
\end{equation}
where
\begin{equation}
H^I_{\ a}=\pi^{JKI}\partial_a\eta_{JK}
-2\sigma_{ab}{}^\|\nabla_cp^{I\,bc}+\sigma_{ab}F^I{}_{Jc}p^{J\,bc}
+P^{IJ}{}_b\partial_aN_J^{\ b}
\end{equation}
The tangential derivative, defined in the appendix, contains a connection coefficient for the normal index and the $F_{IJ}{}^a$ that appears in $H^I_{\ a}$ precisely cancels this coefficient.

When the derivatives are replaced with the momenta the Hamiltonian density becomes
\begin{equation}
H=T+V+N_I^{\ a}H^I_{\ a}
\end{equation}
where $V$ is given by equation (\ref{pot}),
\begin{eqnarray}
T&=&{1\over 2}G_{(IJ)K(LM)N}\pi^{IJK}\pi^{LMN}+G_{(IJ)(KL)}p^I\pi^{JKL}
+{1\over 2}G_{(ab)(cd)}\eta_{IJ}p^{I\,ab}p^{J\,cd}\nonumber\\
&&-{1\over 2}\kappa^2|g|^{-1/2}\sigma_{ab}P_{IJ}{}^aP^{IJb}
\end{eqnarray}
and $p^I=\sigma_{ab}p^{ab}$. The new coefficients are obtained by inverting the matrix
\begin{equation}
\pmatrix{G^{(IJ)K(LM)N}&G^{(IJ)KL}\sigma^{cd}\cr
G^{(LM)NI}\sigma^{ab}&G^{(ab)(cd)}\eta^{IL}\cr}
\end{equation}
The result of the matrix inversion is that
\begin{eqnarray}
G_{IJKLMN} &=&2\kappa^2|g|^{-1/2}\left( 
- {{2 (m - 1)}\over{(d-2)(n - 1)}} 
{\eta}_{IJ} {\eta}_{LM} \eta_{NK}
+ 2 {\eta}_{IL} \eta_{JN}\eta_{KM}\right.
\nonumber \\ 
& & \left.- {2 \over{(n - 1)}} 
\left(\eta_{IL}\eta_{JK}\eta_{MN}+{\eta}_{LM}\eta_{IN}\eta_{JK} 
+ {\eta}_{IJ} \eta_{KL}\eta_{MN}\right)\right)\\
G_{IJKL} &=& 2\kappa^2|g|^{-1/2}\left({2 \over{(d - 2)}} 
{\eta}_{JK} \eta_{IL}\right),
\\
G_{abcd} &=& -2\kappa^2|g|^{-1/2}
\left(2 \sigma_{ac}\sigma_{bd}-{2\over{(d - 2)}} 
\sigma_{ab}\sigma_{cd}\right).\label{gcoef}
\end{eqnarray}
The explicit expressions show that the inversion is possible for codimension $n\ne 1$. When $n=1$, there are $m$ constraints $\pi^{IJK}\equiv P^{IJ}{}_a\equiv0$

The Hamiltonian allows us to construct the action in Hamiltonian form 
\begin{equation}
S = \int d^n x 
\left(\pi^{IJK} {\partial}_K {\eta}_{IJ} 
+ p^{I\,ab} {\partial}_I \sigma_{ab} 
+ P^{IJ}{}_a \partial_I N_J{}^a 
- H \right).\label{action}
\end{equation}
The extrema of this action with respect to independent variations of the momenta $\pi^{IJK}$, $p^{I\,ab}$, $P^{IJ}{}_a$ 
and the fields $\sigma_{ab}$ , $\eta_{IJ}$, $N_I{}^a$ 
gives a first order set of equations. The only constraints on the momenta are the index symmetries, 
\begin{equation}
\pi^{[IJ]K}=p^{I[ab]}=P^{(IJ)}{}_a=0
\end{equation}
If we vary the momenta $\pi^{IJK}$, $p^{I\,ab}$, $P^{IJ}{}_a$ then we find that
\begin{eqnarray}
D_K\eta_{IJ}&=&G_{(IJ)K(LM)N}\pi^{LMN}+G_{(IJ)(KL)}p^L\\
D_I\sigma_{ab}&=&\eta_{IJ}G_{(ab)(cd)}p^{J\,cd}+G_{(LM)(NI)}\pi^{LMN}\\
F^{IJ}{}_a&=&2\kappa^2|g|^{-1/2}P^{IJ}{}_a
\end{eqnarray}
The covariant derivative appears here because the variation of the $H^I_{\ a}$ term in the Hamiltonian provides the necessary connection components. These equations are equivalent to equations (\ref{moma}-\ref{momc}) which define the momenta.

If we vary the action with respect to the fields and substitute for the momenta we obtain exactly the same equations as those obtained from the Lagrangian. Therefore the extrema of the action give a first order form for the Einstein equations decomposed in $n$ independent directions. This is also true in the degenerate case $n=1$, provided that we set $\pi^{IJK}\equiv P^{IJ}{}_a\equiv0$. 

\section{${\rm m}+2$ REDUCTIONS}

When the space of normals to the surface is two dimensional then some simplification occurs. It even becomes possible to fix some of the gauge freedom and still obtain the full set of Einstein equations from the Hamiltonian.

We replace $\eta_{IJ}$ by new variables $(\eta,\tilde\eta_{IJ})$, where
\begin{eqnarray}
\eta&=&(1/2)\log|\eta|\\
\tilde\eta_{IJ}&=&|\eta|^{-1/2}\eta_{IJ}
\end{eqnarray}
This is a canonical transformation if we take conjugate momenta
\begin{eqnarray}
\pi^I&=&\eta_{JK}\pi^{JKI}\\
\tilde\pi^{IJK}&=&
|\eta|^{1/2}\left(\pi^{IJK}-\case1/2\eta^{IJ}\pi^K\right)
\label{tilde}
\end{eqnarray}
The only term that depends on $\tilde \pi^{IJK}$ in the Hamiltonian has the form
\begin{equation}
{1\over 2}G_{IJKLMN}\hat\pi^{IJK}\hat\pi^{LMN}
\end{equation}
where
\begin{equation}
\hat\pi^{IJK}=|\eta|^{-1/2}\tilde\pi^{IJK}-\case1/2(\eta^{IJ}\pi^K
-\eta^{IK}\pi^J-\eta^{JK}\pi^I).
\end{equation}
Variation of $\tilde \pi^{IJK}$ leads to
\begin{equation}
D_K\tilde\eta_{IJ}=|\eta|^{-1/2} G_{IJKLMN}\hat\pi^{LMN}
\end{equation}
If we choose a gauge in which $\tilde\eta_{IJ}$ is constant initially, then $\hat \pi^{IJK}=0$, or equivalently
\begin{equation}
\tilde\pi^{IJK}=\case1/2|\eta|^{1/2}(\eta^{IJ}\pi^K
-\eta^{IK}\pi^J-\eta^{JK}\pi^I).\label{cong}
\end{equation}
Variation of $\tilde\eta_{IJ}$ can be used to show that $D_K\hat\pi^{IJK}=0$, and $\tilde\eta_{IJ}$ remains constant. Variations of the terms involving $\partial_a\tilde\eta_{IJ}$ and $\hat\pi^{IJK}$ vanish and we can drop these terms from the Hamiltonian. 

The Hamiltonian becomes
\begin{equation}
H=T+V+N_I^{\ a}H^I_{\ a}
\end{equation}
where
\begin{eqnarray}
T&=&2\kappa^2|g|^{-1/2}\eta_{IJ}\left(
{1-m\over m}\,\pi^I\pi^J+{2\over m}p^I\pi^J
-\sigma_{ac}\sigma_{bd}p^{I\,ab}p^{J\,cd}+{1\over m}p^Ip^J\right)\nonumber\\
&&-{1\over 2}\kappa^2|g|^{-1/2}\sigma^{ab}
\eta_{IK}\eta_{JL}P^{IJ}{}_aP^{KL}{}_b\\
V&=&-{1\over 4\kappa^2}|g|^{1/2}\sigma^{ab}
\left(\partial_a\eta\,\partial_b\eta+2{}^\|R_{ab}\right)
\end{eqnarray}
and we set $|g|=|\sigma|\exp(2\eta)$ and $\eta_{IJ}=\tilde\eta_{IJ}\exp(\eta)$.

Variation of the momenta gives the equations
\begin{eqnarray}
D_I\eta&=&4\kappa^2|g|^{-1/2}\eta_{IJ}\left({1\over m}p^J+{1-m\over m}\pi^J\right)\label{deta}\\
D_I \sigma_{ab}&=&-4\kappa^2|g|^{-1/2}\eta_{IJ}\left(
p^J{}_{ab}-{1\over m}\sigma_{ab}\,p^J-
{1\over m}\sigma_{ab}\,\pi^J\right)\label{dsigma}\\
F^{IJ}{}_a&=&2\kappa^2|g|^{-1/2}P^{IJ}{}_a\label{dn}
\end{eqnarray}
where $F_{IJ}{}^a$ depends on derivatives of $N_I^{\ a}$ as defined by equation (\ref{f}). Normal indices are raised by $\eta^{IJ}$.

Variation of the fields gives the equations
\begin{eqnarray}
D_I\pi^I&=&{1\over 2\kappa^2}|g|^{1/2}\left({}^\|R-
{}^\|\nabla^2\eta-{1\over 2}({}^\|\nabla\eta)^2+
{1\over4}F_{IJa}F^{IJa}\right)\label{dpi}\\
D_I( \sigma_{ac}p^{I\,bc})&=&
{1\over 4\kappa^2}\sigma_a^{\ b}\eta_{IJ}\left(
{1-m\over m}\pi^I\pi^J+{2\over m}p^I\pi^J-p^{Icd}p^J_{\ cd}+
{1\over m}p^Ip^J\right)\nonumber\\
&&+{1\over 2\kappa^2}|g|^{1/2}
\left({}^\|\nabla_a{}^\|\nabla^b\eta-\sigma_a^{\ b}{}^\|\nabla^2\eta
+{1\over 2}{}^\|\nabla_a\eta{}^\|\nabla^b\eta-
{3\over 4}\sigma_a^{\ b}({}^\|\nabla\eta)^2\right)\nonumber\\
&&-{1\over 8\kappa^2}|g|^{1/2}
\left(F_{IJ}{}_aF^{IJb}+
{1\over 2}\sigma_a^{\ b}F_{IJ}{}_cF^{IJc}\right)\nonumber\\
&&-{1\over 2\kappa^2}|g|^{1/2}
\left({}^\|R_a^{\ b}-{1\over 2}{}^\|R\sigma_a^{\ b}\right)
\label{dp}\\
D_JP^{IJ}{}_a&=&\pi^I\partial_a\eta-2\sigma_{ab}{}^\|\nabla_cp^{I\,bc}
+\sigma_{ab}F^I{}_{Jc}p^{J\,bc}
\label{mc}
\end{eqnarray}
These equations are far simpler than those derived from the action (\ref{action}).

These equations can be put into second order form by eliminating the momenta and producing equations for the metric components. Expression for the momenta follow from equations (\ref{deta}-\ref{dn}),
\begin{eqnarray}
p^{I\,ab}&=&-{1\over 2\kappa^2}|g|^{1/2}
\left(k^{I\,ab}-\sigma^{ab}k^I+{1\over 2}\sigma^{ab}D^I\eta\right)\\
\pi^I&=&{1\over 2\kappa^2}|g|^{1/2}k^I
\end{eqnarray}
where the extrinsic curvature $k_{Iab}=(1/2)D_I\sigma_{ab}$. If we differentiate equations (\ref{dsigma}) and (\ref{deta}),
\begin{eqnarray}
D_I|g|^{1/2}\sigma^{bc}D^I\sigma_{ac}&=&
-2\left(D_I\sigma_{ac}p^{I\,bc}\right)_{TF}+
{2\over m}\sigma_a^{\ b}D_I\pi^I\\
D_I|g|^{1/2}D^I\eta&=&
{2\over m}D_Ip^I+\left({2\over m}-2\right)D_I\pi^I
\end{eqnarray}
where $()_{TF}$ denotes the traceless part of the tensor. Now we use equations (\ref{dp}) and (\ref{dpi}) to get
\begin{eqnarray}
|g|^{-1/2}D_I\left(|g|^{1/2}k^{I\ \,b}_{\ a}\right)&=&
{}^\|R_a^{\ b}-{}^\|\nabla_a{}^\|\nabla^b\eta-
{1\over2}{}^\|\nabla_a\eta{}^\|\nabla^b\eta+{1\over 4}F_{IJ\,a}F^{IJ\,b}\label{e1}\\
|\eta|^{-1/2}D_I\left(|\eta|^{1/2}D^I\eta\right)&=&
-{}^\|R-{1\over 2}({}^\|\nabla\eta)^2-{1\over 4}F_{IJ\,a}F^{IJ\,a}
-\left(k_{I\,ab}k^{I\,ab}-k^Ik_I\right)\label{e2}
\end{eqnarray} 
In addition to these equations we still have equation (\ref{mc}),
\begin{equation}
|\sigma|^{-1/2}D_J\left(|\sigma|^{1/2}F_I^{\ J}{}_a\right)=
(2\eta_I^{\ J}{}^\|\nabla_b-F_{I\ \,b}^{\ J})(k^{\ \ b}_{Ja}
-\sigma^{\ b}_ak_J-\case1/2\sigma^{\ b}_aD_J\eta)
+k_I\partial_a\eta\label{e3}
\end{equation}
Equations (\ref{e1}--\ref{e3}) are a complete set of vacuum Einstein equations and are equivalent to the vanishing of the Ricci tensor. They are second order in derivatives of the metric when used in conjunction with equations (\ref{kld}) for $k^I_{\ ab}$ and (\ref{feq}) for $F_{IJ}{}^a$.

A similar set of equations, for the $2+2$ decomposition with null coordinates, has been given by Brady et al. \cite{brady}. In their case
\begin{equation}
\eta_{IJ}=e^\eta\pmatrix{0&-1\cr-1&0\cr}
\end{equation}
With the change of notation described in Table 1 of the appendix, and some trivial rearrangements, the two sets of equations agree. One advantage of our more general formalism is that we have been able to derive all of the their equations from an action principal.   

Another important example is when the submanifolds have $m$ commuting killing vectors and the metric decomposition is orthogonal, $F_{IJ}{}^a=0$. In this case, equation (\ref{e3}) vanishes identically and the equations separate, making it possible to solve (\ref{e1}) for $\sigma_{ab}$. The solution can then be substituted into equation (\ref{e2}) to obtain $\eta$. This special case includes the Kerr metric and its generalisations.

\section{BOUNDARY CONTRIBUTIONS TO THE HAMILTONIAN}

We turn now to the application of the canonical formalism to manifolds with boundaries, where the metric on the boundary is specified and the boundary terms that were discarded earlier will have to be included. This can be readily adapted to the situation in which only the intrinsic metric on the boundary is fixed, and we shall see how the action can be recovered.

Problems of this type are often encountered in the path integral approach to quantum gravity. The $3+1$ formalism can be used to decompose the action for manifolds with boundary and it has been analysed in this context by York\cite{york3} and Hawking and Hunter\cite{hawking2}. We will see how the generalised canonical approach leads to a new understanding of the $3+1$ case and some improvements on the previous results. Because the results have applications to quantum gravity, we will make explicit use of Riemannian signature $(++++)$ in this section.

The decomposition of the action lead to two total divergences, equations (\ref{b1}) and (\ref{b2}). We shall add a boundary term to the original action to cancel these divergences
\begin{equation}
S^{m+n}={1\over2\kappa^2}\int_{\cal M}d\mu R+
{1\over 2\kappa^2}\int_{\partial \cal M}
d\omega_I\left(2k^I-\eta^{IL}\eta^{JK}
(D_J\eta_{LK}-D_J\eta_{IK})\right)\label{newaction}
\end{equation}
This assumes that the boundary is given by a function of the $x^I$ only. We set  $d\omega_I=n_I^{\ a}d\omega_a$, where ${\bf d}\mbox{\boldmath$\omega$}$ is the surface form and $d\mu$ is the volume measure.

After canonical decomposition the Lagrangian is first order in field derivatives. Variation of the action gives the field equations, plus a boundary term 
\begin{equation}
(\delta S^{m+n})_{\partial\cal M}=\int_{\partial\cal M}|g|^{-1/2}d\omega_I
\left(\pi^{IJK}\delta\eta_{JK}+p^{I\,ab}\delta \sigma_{ab}
+P^{IJ}{}_a\delta N_J^{\ a}\right).\label{vars}
\end{equation}
The field equations can be deduced from the variational principal if the metric is fixed on the boundary. We can also take the value of the action for solutions to the field equations, calling this the principal function, then equation (\ref{vars}) relates the momenta to the variation of the principal function.

\begin{figure}
\begin{center}
\leavevmode
\epsfxsize=20pc
\epsffile{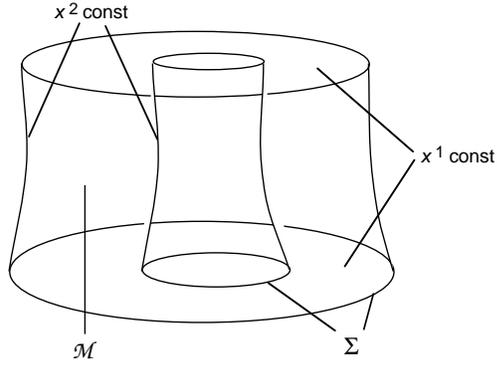}
\end{center}
\caption{The manifold ${\cal M}$ with a boundary made up from two surfaces of constant $x^1$ and two surfaces of constant $x^2$ intersecting in four corners of dimension two. \label{fig1}}
\end{figure}

The action $S^{m+n}$ is suitable for fixing all of the metric components on the boundary but it is not yet suitable for variations which fix only the intrinsic components of the metric. We will obtain the necessary modifications to the action for the boundary shown in figure \ref{fig1} and compare this with known results.

\begin{figure}
\begin{center}
\leavevmode
\epsfxsize=20pc
\epsffile{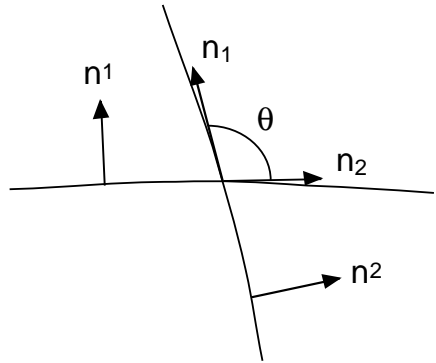}
\end{center}
\caption{Surfaces of constant $x^I$, $I=1,2$, with the basis of normal vectors ${\bf n}^I$ and the dual basis ${\bf n}_I$. The angle between the dual basis vectors is $\theta$\label{fig2}}
\end{figure}

We assume that the space can be foliated by two sets of coordinate surfaces which include the boundary, allowing a $2+2$ decomposition. Figure \ref{fig2} shows part of the boundary, where ${\bf n}^I$ are normal forms and the dual vectors ${\bf n}_I$ are tangential to the boundary surfaces. We let $\theta$ be the angle between the tangential vectors, then $\theta$ can be related to the metric $\eta_{IJ}$ by
\begin{eqnarray}
\eta_{12}&=&(\eta_{11}\eta_{22})^{1/2}\cos\theta\label{defth}\\
|\eta|^{1/2}&=&(\eta_{11}\eta_{22})^{1/2}\sin\theta
\end{eqnarray}
We also make use of the traced extrinsic curvature of the boundary $K^I=\nabla\cdot{\bf n}^I$. Comparison with the definition (\ref{defk}) gives
\begin{equation}
K^I=k^I+c^{JI}{}_J-(\eta^{II})^{-1}c^{III}\label{defki}
\end{equation}
with no sum over $I$.
   
With a $2+2$ decomposition it is possible to take $\eta_{IJ}$ to be conformally constant, and by equation (\ref{cong})
\begin{equation} 
\pi^{IJK}=\eta^{IJ}\pi^K-\case1/2\eta^{JK}\pi^I-\case1/2\eta^{IK}\pi^J
\end{equation}
This simplifies equation (\ref{vars}),
\begin{equation}
\pi^{IJK}\delta\eta_{IJ}=(\eta^{IJ}\pi^K-\eta^{JK}\pi^I)\delta\eta_{IJ}
\end{equation}
The antisymmetry in the indices $I$ and $K$ allows us to introduce the antisymmetric tensor $\epsilon_{IJ}$,
\begin{equation}
\pi^{IJK}\delta\eta_{IJ}=\epsilon^{IK}\epsilon^{JL}\pi_L\delta\eta_{IJ}
\end{equation}
The momenta can also be replaced by the extrinsic curvatures
\begin{equation}
\pi^{IJK}\delta\eta_{IJ}={1\over2\kappa^2}|g|^{1/2}
\epsilon^{IK}\epsilon^{JL}k_L\delta\eta_{IJ}
\end{equation}
Since this vanishes when $I=K$, the variation of the action does not depend upon the purely normal metric components but it does depend upon the mixed components $\delta\eta_{12}$. We can eliminate these by using a new action $S^{\rm cov}$, obtained by adding an extra surface term.

Only one extra surface term, depending on $\theta$ and $k_I$, is needed. The surface term becomes (compare equation (\ref{newaction}))
\begin{equation}
{1\over 2\kappa^2}\int_{\partial \cal M}
d\omega_I\left(2k^I-\eta^{IL}\eta^{JK}
(D_J\eta_{LK}-D_J\eta_{IK})-\theta\epsilon^{IJ}k_J\right)
\end{equation}
Variation of the final term cancels the $\delta\eta_{12}$. 

We can rewrite the boundary term in terms of the extrinsic curvature $K^I$ of the boundary using the definition (\ref{defki}). We also use the expression (\ref{expc}) for $c_{IJK}$, to get a familiar expression
\begin{equation}
S^{\rm cov}={1\over2\kappa^2}\int_{\cal M}d\mu R+
{1\over \kappa^2}\int_{\partial \cal M}
d\omega_I\,K^I+
{1\over \kappa^2}\int_{\Sigma}d\mu
(\theta_0-\theta)\label{hs}
\end{equation}
where $\Sigma$ denotes the corners of the boundary, $d\mu$ the volume measure and $\theta_0$ is a constant. For an additive action we take $\theta_0=\pi/2$.

The corner terms were first discovered in the context of Regge calculus by Hartle and Sorkin\cite{hartle}. They were rediscovered later with Lorentzian signature\cite{hayward1,hayward2}. In the Lorenzian case we replace $\theta_0-\theta$ by $i\beta$, where $\beta$ is a Lorentz boost. 

\begin{figure}
\begin{center}
\leavevmode
\epsfxsize=20pc
\epsffile{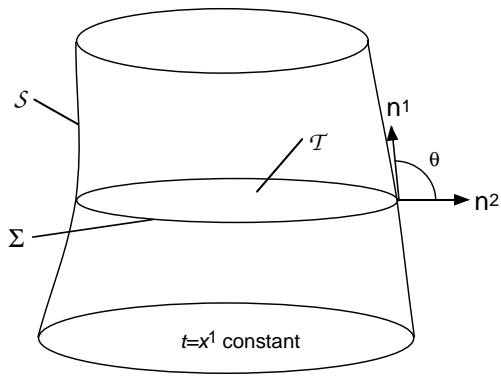}
\end{center}
\caption{The manifold ${\cal M}$ with boundary ${\cal S}$ and two surfaces of constant time.\label{fig3}}
\end{figure}

For the remainder of this section we will decompose $S^{\rm cov}$ into $3+1$ form and construct the corresponding Hamiltonian. We will take $x^1$ to be the time coordinate, running over a fixed interval, and $x^2$ a radial coordinate, running from zero to a fixed value $x^2_s$ on the edge ${\cal S}$. The coordinates on the surfaces of constant time will be denoted by $x^i$, and
\begin{equation}
x^2(x^i,t)=x^2_s
\end{equation}
on ${\cal S}$, as shown in figure (\ref{fig3}). The metric becomes 
\begin{equation}
{\bf g}=N^2{\bf dt}\otimes{\bf dt}+
h_{ij}({\bf dx}^i+N^i{\bf dt})\otimes({\bf dx}^j+N^j{\bf dt}).
\end{equation}
The $3+1$ and the $2+2$ form of the metric are related by $\eta^{11}=N^{-2}$ and $N_{1a}=h_{ai}N^i$.

The first step is to construct the Lagrangian. We separate the volume terms from the boundary terms and write $S^{\rm cov}=S_{\cal M}+S_{\partial\cal M}$. The $3+1$ decomposition of the curvature introduces divergences $\nabla\cdot{\bf v}$, with ${\bf v}$ given by of equation (\ref{b1}) in codimension 1. This introduces boundary terms of the form ${\bf n}^I\cdot{\bf v}$, which become 
\begin{eqnarray}
{\bf n}^1\cdot{\bf v}&=&-2K^1\\
{\bf n}^2\cdot{\bf v}&=&2(\eta^{11})^{-1}
\left((\eta_{22})^{-1}c^{11}{}_2-K^1\eta^{12}\right)
\end{eqnarray}
Boundary terms on the surfaces of constant time cancel with boundary terms already present in the action $S^{\rm cov}$ (equation \ref{hs}). Combining terms on the boundary $\cal S$ with $S^{\rm cov}$ gives
\begin{equation}
S_{\partial\cal M}={1\over\kappa^2}
\int_{\cal S}d\omega_2(\eta^{11})^{-1}
\left((\eta_{22})^{-1}c^{11}{}_2-K^1\eta^{12}
+\eta^{11}K^2\right)
\end{equation}
There are also corner terms. Next we replace the extrinsic curvatures using the identity
\begin{equation}
k_2=\eta_{2I}K^I-\eta^{11}|\eta|^{1/2}D_1\theta+
(\eta_{22})^{-1}|\eta|c^{11}{}_2,
\end{equation}
which can be derived from from equation (\ref{defki}) and (\ref{expc}). We are left with
\begin{equation}
S_{\partial\cal M}={1\over\kappa^2}
\int_{\cal S}d\mu dt\left((\eta^{11}\eta_{22})^{-1/2}k_2
-(\theta-\theta_0)k_1\right)\label{cans}
\end{equation}
When $\theta_0=\pi/2$, the corner terms cancel and this expression is the complete expression for boundary contributions to the Lagrangian.

The surface term (\ref{cans}) has two components. Appart from at the corners, the value of $k_1$ is fixed by the boundary conditions. However, $k_1$ includes time derivatives of the metric,
\begin{equation}
k_1=\case1/2\sigma^{ab}\left(\partial_t\sigma_{ab}-2{}^\|\nabla_aN_{1b}\right)
\end{equation}
When we take the functional derivative of the the Lagrangian with respect to $\partial_t\sigma_{ab}$ we get surface momenta,
\begin{equation}
\hat p^{ab}={1\over 2\kappa^2}|\sigma|^{1/2}(\theta_0-\theta)\sigma^{ab}\label{surfmom}
\end{equation}
This suggests that the angle $\theta$ should be regarded as a momentum variable. Since the value of $\theta$ on the corners depends upon the normal metric components it is not fixed by the boundary conditions.

In canonical form, the surface terms (\ref{cans}) become
\begin{equation}
S_{\partial\cal M}=
\int_{\cal S}d^2x dt\left(
\hat p^{ab}\partial_t\sigma_{ab}+
N\kappa^{-2}(\eta_{22})^{-1/2}|\sigma|^{1/2}k_2+
N^ih_{ia}{}^\|\nabla_b\hat p^{ab}\right)\label{sbo}
\end{equation}
The usual ADM Hamiltonian can be obtained by specialising the earlier discussion to $n=1$, setting $\pi^{IJK}\equiv 0$. This introduces an additional surface term from a divergence
\begin{equation}
-2{}^{(3)}\nabla_j(N_ip^{ij})\label{haml}
\end{equation}
Combining the surface terms together gives
\begin{equation}
\int_{\cal S}d^2x dt\left(
\hat p^{ab}\partial_t\sigma_{ab}-
N{\cal H}-N^i{\cal H}_i\right)
\end{equation}
The surface contributions to the Hamiltonian are,
\begin{eqnarray}
{\cal H}&=&-\kappa^{-2}(\eta_{22})^{-1/2}D_2|\sigma|^{1/2}\\
{\cal H}_i&=&-h_{ia}{}^\|\nabla_b\hat p^{ab}
-(\eta_{22})^{-1/2}h_{ij}n_{2k}p^{jk}
\end{eqnarray}
Variation of the full action, subject to the condition (\ref{surfmom}), gives the field equations with no surface terms.
 
The term ${\cal H}_i$ looks very different from the results of Hawking and Hunter \cite{hawking2} because we have separated the momentum term in (\ref{sbo}). On the other hand, the value of the action is identical in the two cases. 

We can further verify the existence of surface momenta by direct variation of the action $S^{\rm cov}$. This gives the field equations plus surface terms
\begin{equation}
\delta S^{\rm cov}=\int_{\partial\cal M}d^3x
\,p^{ij}\delta h_{ij}+\int_{\Sigma}d^2x\,\hat p^{ab}\delta\sigma_{ab}.\label{var}
\end{equation}
The value of $S^{\rm cov}$ for solutions to the field equations gives Hamilton's principal function. This will be a function of the metric on $\partial\cal M$. Equation (\ref{vars}) shows that the principal function has a variation with support on the corners. Hamilton-Jacobi theory tells us that this variation has to be a momentum. It also follows that one family of constants of integration of the Hamilton--Jacobi equation will be the initial values of the surface momenta, or equivalently, the initial values of $\theta$.

\section{CONCLUSION}

We have described a generalisation of the Hamiltonian formalism for gravity that is appropriate to foliations of codimension $n>1$. The new formalism puts the vacuum Einstein equations into a first-order form. It also has a variational formulation which allows canonical transformations to be used to simplify the classical field equations.

The results presented here can be extended to include matter fields. We hope that they will be valuable for finding new solutions to Einstein-matter systems in various spacetime dimensions.    

\appendix
\section*{GENERAL FORMALISM}

Suppose we have a metric ${\bf g}$ on a $d$-dimensional spacetime ${\cal M}$ and a sub-surface $\Sigma$ of dimension $m < d$. Using a set of normal forms ${\bf n}^I$ and tangential forms $\mbox{\boldmath$\omega$}^a$ we can put the metric in the form
\begin{equation}
{\bf g} =  \sigma_{ab}\mbox{\boldmath$\omega$}^a \otimes \mbox{\boldmath$\omega$}^b +
{\eta}_{IJ} {\bf n}^I \otimes {\bf n}^J,
\end{equation}
The Levi-Cevita connection $\nabla$ on ${\cal M}$ induces a connection on the tangent space to $\Sigma$ and we can obtain Gauss--Codazzi relations for the curvature. There does not exist a standard notation for these reductions and so we set out ours here.

The inverse metric
\begin{equation}
\tilde{\bf g} = \sigma^{ab} {\bf e}_a \otimes {\bf e}_b+
{\eta}^{IJ} {\bf n}_I \otimes {\bf n}_J,
\end{equation}
where the vectors ${\bf e}_a$ and ${\bf n}_I$ are dual to the forms $\mbox{\boldmath$\omega$}^a$ and ${\bf n}^I$. We call tensors with components along ${\bf e}_a$ or $\mbox{\boldmath$\omega$}^a$ tangential and those with components along ${\bf n}_I$ and ${\bf n}^I$ normal.

We define the extrinsic curvature and related quantities,
\begin{eqnarray}
k^I{}_{ab} &=& 
\sigma_a{}^{a'} \sigma_b{}^{b'} (\nabla_{a'}{\bf n}^I)_{b'},\label{defk}
\\
a^{IJ}{}_a &=& 
\sigma_a{}^{a'} n^{Ib'}(\nabla_{a'}{\bf n}^I)_{b'} ,
\\
b^{IJ}{}_a &=& 
\sigma_a{}^{a'} n^{Ib'}(\nabla_{a'}{\bf n}^I)_{b'} ,
\\
c^{IJK} &=& n^{Ka'}n^{Ib'}(\nabla_{a'}{\bf n}^I)_{b'}.\label{defc} 
\end{eqnarray}
These transform like connections under changes in the normal forms, but they can be regarded as rank 3 tensors if the basis is changed whilst keeping the normal forms fixed. 

The hypersurface orthogonality of the ${\bf n}^I$ implies that the extrinsic curvature is symmetric in the tangential indices. The tensors $b^{IJ}_{\ \ a}$ and $c^{IJK}$ are identically zero in the most familiar situation where the surface has codimension 1 and the normals have unit length. In the general case they can be interpreted as connection coefficients in the normal directions. This can be seen in the Gauss--Weingarten equations that follow from the definitions given above,
\begin{eqnarray}
\nabla{\bf n}^I&=&{\bf k}^I+a_J{}^I{}_b{\bf n}^J\otimes{}\mbox{\boldmath$\omega$}^b
+b_J^{\ I}{}_a\mbox{\boldmath$\omega$}^a\otimes{\bf n}^J+
c_{J\ K}^{\ I}{\bf n}^J\otimes{\bf n}^K\\
\nabla{\bf n}_I&=&{\bf k}_I+a_{JI}{}^b{\bf n}^J\otimes{\bf e}_b
-b_I^{\ J}{}_a\mbox{\boldmath$\omega$}^a\otimes{\bf n}_J-
c_{J\ I}^{\ K}{\bf n}^J\otimes{\bf n}_K\label{gw}
\end{eqnarray}

We can now decompose the connection in directions tangential and normal to the surface,
\begin{eqnarray}
{}^\|\nabla_a T^{J\dots K}{}_{L\dots M}{}^{b\dots c}{}_{d\dots e}&=&
\nabla_{a'}T^{J\dots K}{}_{L\dots M}{}^{b\dots c}{}_{d\dots e}\\
{}^\perp\nabla_I T^{J\dots K}{}_{L\dots M}{}^{b\dots c}{}_{d\dots e}&=&
n_I^{\ a'}
\nabla_{a'}T^{J\dots K}{}_{L\dots M}{}^{b\dots c}{}_{d\dots e}\label{perpd}
\end{eqnarray}
The derivatives have been projected so that tangential indices remain tangential and normal indices remain normal. These decomposed derivatives are both metric connections
\begin{equation}
{}^\|\nabla_a \sigma_{bc}={}^\perp\nabla_I\eta_{JK}=0.
\end{equation}
However, the perpendicular connection has torsion $a^{[IJ]}{}_a$.
 
With these derivatives, we can now complete the Gauss--Weingarten equations and deduce that
\begin{eqnarray}
\nabla{\bf n}_I&=&{}^\|\nabla{\bf n}_I+{}^\perp\nabla{\bf n}_I+
k_{Ia}{}^b\mbox{\boldmath$\omega$}^a\otimes{\bf e}_b+
a_{JI}{}^b{\bf n}^J\otimes{\bf e}_b\\
\nabla{\bf e}_b&=&{}^\|\nabla{\bf e}_b+{}^\perp\nabla{\bf e}_b
-k^I_{\ ab}\mbox{\boldmath$\omega$}^a\otimes{\bf n}_I-
a_{I\ b}^{\ J}{\bf n}^I\otimes{\bf n}_J
\end{eqnarray}
It is often convenient to replace ${}^\perp\nabla$ and use instead the tangential projection of the Lie derivative along ${\bf n}_I$, which we denote by $D_I$. The vanishing torsion of the Levi-Cevita connection $\nabla$ implies that the two are related by
\begin{eqnarray}
{}^\perp\nabla{\bf e}_b&=&D{\bf e}_b+
k_{Ib}{}^c{\bf n}^I\otimes{\bf e}_c\\
{}^\perp\nabla\mbox{\boldmath$\omega$}^b&=&D\mbox{\boldmath$\omega$}^b-
k_{I\ c}^{\ b}{\bf n}^I\otimes\mbox{\boldmath$\omega$}^c\label{dconv}
\end{eqnarray}

The curvature tensors of the connections will be denoted by ${}^\|{\bf R}$ and ${}^\perp{\bf R}$ respectively. The perpendicular derivative of a normal vector ${\bf X}$ is given by
\begin{equation}
{}^\perp\nabla_IX^J={\bf n}_I(X^J)-c_{I\ K}^{\ J}X^K
\end{equation}
The commutator of two of these derivatives gives an expression for ${}^\perp{\bf R}$ in terms of the Gauss--Wiengarten coefficients,
\begin{equation}
{}^\perp R^{IJ}{}_{KL}=2{}^\perp\nabla{}_{[L} c{}_{K]}{}^I{}_J
+2c_{[K}^{\ \,MJ}c_{L]\ M}^{\ \,I}+2b^{JI}{}_aa_{[KL]}{}^a
\end{equation}
The final term, that depends on the torsion, comes from the commutator $[{\bf n}_K,{\bf n}_L]$. Similarly, we also have an expression for the commutator of two tangential derivatives ${}^\|\nabla$ acting on normal vectors,
\begin{equation}
{}^\| R^{IJ}{}_{ab}=2{}^\|\nabla{}_{[a}b^{IJ}{}_{b]}-
b^{IK}{}_ab^J{}_{Kb}+b^{IK}{}_bb^J{}_{Ka}
\end{equation}

Taking commutators of $\nabla$ on the Gauss--Weingarten relations we find Gauss-Codazzi relations for the tangential $a,b\dots$ and normal components $I,J\dots$ of the curvature tensor ${\bf R}$ of the connection $\nabla$,
\begin{eqnarray}
R^a_{\ bcd}
&=&{}^\| R^a_{\ bcd} - {\eta}_{IJ} 
\left( k^I{}^a{}_c k^{J}{}_{bd} - k^I{}^a{}_d k^{J}{}_{bc} \right)\\
R_{a\ \,bc}^{\ I}&=&{}^\|\nabla_bk^I{}_{ac}-{}^\|\nabla_ck^I{}_{ab}\\
R^{I\ \,J}_{\ a\ \,b} &=&
-{}^\perp\nabla^Jk^I_{\ ab}-k^I_{\ ac}k^{J\,c}{}_b+
{}^\|\nabla_ba^{JI}{}_a-a_{K\ \,a}^{\ \ I}a^{JK}{}_b.\\
R^{IJ}{}_{ab}&=&{}^\|R^{IJ}{}_{ab}-
k^I{}^c{}_a k^{J}{}_{bc} + k^I{}^c{}_b k^{J}{}_{ac}\\
R^a{}_{IJK}&=&{}^\perp\nabla_Ja_{KI}{}^a-{}^\perp\nabla_Ka_{JI}{}^a
-2k_{I\ b}^{\ a}a_{[JK]}{}^b\\
R^I_{\ \,JKL}&=&{}^\perp R^I_{\ \,JKL}-
a_{K\ \,b}^{\ \ I}a_{LJ}{}^b+a_L^{\ \ I}{}_ba_{KJ}{}^b.
\end{eqnarray}
In our notation, the covariant derivatives include connection coefficients for the normal indices. 

By taking traces of the above relations, we have a decomposition of the Ricci scalar $R$:
\begin{eqnarray}
R&=&{}^\|R+{}^\perp R+k_Ik^I-k_{Iab}k^{Iab}\nonumber\\
&&+a^I_{\ I}{}_aa^J_{\ J}{}^a-a^{IJa}a_{JIa}
+\nabla_av^a.\label{ricci}
\end{eqnarray}
where
\begin{equation}
v^a=-2n_I^{\ \,a}k^I+2a^I_{\ I}{}^a.
\end{equation}

A special case of interest is when the manifold ${\cal M}$ has a foliation of codimension $n$. In this case we can choose a coordinate basis,
\begin{equation}
{\bf n}^I = dx^I, \qquad \mbox{\boldmath$\omega$}^a = dx^a + N_I{}^a dx^I.
\end{equation}
The dual basis
\begin{equation}
{\bf n}_I = {\partial}_I - N_I{}^a {\partial}_a,\qquad
{\bf e}_a = {\partial}_a,\label{basis}
\end{equation}
where ${\partial}_I =\partial/\partial x^I$ and ${\partial}_a =\partial/\partial x^a$.

In this basis the Gauss--Weingarten coefficients are related by $a^{IJ}{}_a=b^{IJ}{}_a$ and $c^{IJK}=c^{KIJ}$. Because of these relationships the derivative $D_I$ satisfies
\begin{equation}
D_I{\bf n}_I=D_I{\bf n}^I=0.
\end{equation}
Therefore there are no connection terms for the normal indices. Furthermore, the tensor $F_{IJ}$, defined by
\begin{equation}
F_{IJ}=[D_I,D_J]
\end{equation}
can also be written
\begin{equation}
F_{IJ}{}^a={\partial\strut}_JN_I^{\ a}-{\partial\strut}_IN_J^{\ a}-
N_J^{\ b}\partial_bN_I^{\ a}+N_I^{\ b}\partial_bN_J^{\ a}\label{feq}
\end{equation}
It depends only on the $N_I^{\ a}$.

Now we can express the Gauss--Weingarten coefficients in terms of the derivative $D_I$. First of all, using ${}^\perp\nabla\sigma_{ab}=0$ and equation (\ref{dconv}), we find that  
\begin{equation}
k^I_{\ ab}=\case1/2\eta^{IJ}D_J\sigma_{ab}\label{kld}
\end{equation}
Writing out the Lie derivative in the coordinate basis gives,  
\begin{equation}
D_I\sigma_{ab}={\bf n}_I(\sigma_{ab})-\sigma_{ac}\partial_bN_I{}^c
-\sigma_{bc}\partial_aN_I{}^c
\end{equation}

In the coordinate basis we get
\begin{equation}
a^{IJ}{}_a = b^{IJ}{}_a =
\case1/2  F^{IJ}{}_a
+\case1/2 {\partial}_a({\eta}^{IJ}).
\end{equation}
Finally, by regarding the $c^{IJK}$ as connection coefficients we may deduce that 
\begin{equation}
c_{IJK}=-\case1/2D_I\eta_{JK}-
\case1/2D_K\eta_{IJ}+\case1/2D_J\eta_{IK}.\label{expc}
\end{equation}
The table gives comparisons between the notation used here and the notation used in two other references.

\begin{table}
\begin{tabular}{lll}
GM&BDIM&HH\\
\hline
\\
$\eta_{IJ}$&$\displaystyle\pmatrix{0&-e^\lambda\cr-e^\lambda&0\cr}$&
$\displaystyle\pmatrix{-1&\sinh\eta\cr\sinh\eta&1\cr}$\\
$\sigma_{ab}$&$g_{ab}$&$\sigma_{ab}$\\
${\bf n}_I$&$\ell_A$&$r_i\,{\rm sech}\eta$\\
${\bf n}^I$&$e^{-\lambda}\ell^A$&$n_\mu$\\
$k^I{}_{ab}$&$K^A{}_{ab}$&$k_{ab}$\\
${}^\|\nabla_a$&$\nabla_a$ (see caption)&\\
$D_I$&$D_A$&\\
$a_{IJ}{}^a$&$L_{AB}{}^a$&\\
$c_{IJK}$&$N_{BCA}$&\\
$F_{IJ}{}^a$&$\epsilon_{BA}\omega^a$&\\
\end{tabular}
\caption{This table compares the notation of the present paper (GM) with reference  (BDIM) and  (HH). The index $a=1\dots m$, where BDIM and HH both have $m=2$. The derivative ${}^\|\nabla_a$ includes connections on normal indices but $\nabla_a$ does not.}
\label{tabb}
\end{table}

\end{document}